\documentclass[preprint,12pt]{elsarticle}

\usepackage{graphicx}
\usepackage{amssymb}
\usepackage{amsmath}
\usepackage{booktabs}
\usepackage{array}
\biboptions{comma,square,sort&compress}

\def\rmd{{\rm d}}
\def\rme{{\rm e}}
\journal{Astroparticle Physics}

\begin{document}

\begin{frontmatter}

\title{On the absolute value of the air-fluorescence yield}


\author{J.~Rosado\corref{cor}}
\ead{jaime\_ros@fis.ucm.es}

\author{F.~Blanco}

\author{F.~Arqueros}

\address{Departamento de F\'{i}sica At\'{o}mica, Molecular y Nuclear, Facultad de Ciencias F\'{i}sicas,
Universidad Complutense de Madrid, E-28040 Madrid, Spain}

\cortext[cor]{Corresponding author}

\begin{abstract}
The absolute value of the air-fluorescence yield is a key parameter for the energy reconstruction of extensive air
showers registered by fluorescence telescopes. In previous publications, we reported a detailed Monte Carlo simulation
of the air-fluorescence generation that allowed the theoretical evaluation of this parameter. This simulation has been
upgraded in the present work. As a result, we determined an updated absolute value of the fluorescence yield of
$7.9\pm2.0$~ph/MeV for the band at 337~nm in dry air at 800~hPa and 293~K, in agreement with experimental values. We
have also performed a critical analysis of available absolute measurements of the fluorescence yield with the
assistance of our simulation. Corrections have been applied to some measurements to account for a bias in the
evaluation of the energy deposition. Possible effects of other experimental aspects have also been discussed. From this
analysis, we determined an average fluorescence yield of $7.04\pm0.24$~ph/MeV at the above conditions.
\end{abstract}


\end{frontmatter}

% \linenumbers

\section{Introduction}
\label{sec:intro}

Ultra-high-energy cosmic rays are efficiently detected from the isotropic fluorescence light produced in the atmosphere
by the extensive air showers. This technique was used by HiRes \cite{HiRes} and is presently employed by the Pierre
Auger Observatory \cite{Auger} and the Telescope Array experiment \cite{TA}. The JEM-EUSO Collaboration \cite{JEM-EUSO}
has also proposed to record the air-fluorescence traces of cosmic rays from the top of the atmosphere in a satellite
mission. The number of fluorescence photons emitted at a given shower position is proportional to the energy deposited
in the atmosphere; hence, the longitudinal development of the light registered by fluorescence telescopes provides a
calorimetric measure of the shower energy. The fluorescence yield (FY) in air, i.e., the number of photons emitted per
unit energy deposited in the atmosphere, is thus a key parameter in this technique.

The production of fluorescence light in the atmosphere has been studied since long ago in various fields, e.g., in
aurora physics \cite{Grun}. This light emission is due to the de-excitation of molecules previously excited by charged
particles.\footnote{Strictly speaking, this radiation should be named scintillation instead of fluorescence.} In the
spectral range of interest in the cosmic-ray field (i.e., around 300--400~nm), the air fluorescence is dominated by the
second positive (2P) band system of N$_2$ (C~$^3\Pi_{\rm u}\rightarrow$~B~$^3\Pi_{\rm g}$) and the first negative (1N)
band system of N$_2^+$ (B~$^2\Sigma^+_{\rm u}\rightarrow$~X~$^2\Sigma^+_{\rm g}$). Excited species can also relax
without emitting light through collisions with other molecules \cite{Stern-Volner}; as a consequence, the fluorescence
intensity depends on the atmospheric conditions (i.e., pressure, temperature and humidity).

Although the processes underlying the generation of this radiation are fully understood since the early twentieth
century, the associated parameters have not been known with enough precision for an accurate energy reconstruction of
air showers until very recently. The most convenient procedure for the analysis of cosmic-ray data is to combine an
accurate value of the absolute FY at given air conditions for a reference molecular band (or a wavelength interval)
with the relative intensities of the fluorescence spectrum and those parameters related to the atmospheric
dependencies. This paper is focused on the absolute value of the FY. Note that the uncertainty in this parameter
translates almost linearly to the energy scale of the fluorescence telescopes.

In previous works \cite{PhysLett,AstropartPhys1,NIMA_th,NJP,AstropartPhys2}, we presented a theoretical evaluation of
the FY based on a Monte Carlo (MC) simulation of both the energy deposition and fluorescence emission. It was shown
that the nitrogen fluorescence is mostly induced by low-energy secondary electrons generated along the tracks of
energetic charged particles. Consequently, our main efforts were directed toward a detailed characterization of the
generation and transport of these secondary electrons. In this paper, several relevant upgrades of our MC algorithm are
described (section \ref{sec:upgrades}) and the corresponding updated results are reported (section
\ref{sec:MC_results}). Details are given in two Appendices.

Several measurements of the absolute FY were carried out in the past years
\cite{Kakimoto,Nagano,Lefeuvre,MACFLY,FLASH,AirLight,Dandl,Airfly}. The experimental technique usually consists of a
beam of electrons that collides with an air target at known conditions and an appropriate detection system that
measures the absolute fluorescence intensity. In \cite{NJP,AstropartPhys2}, we compared the available FY data
normalized to common conditions and discussed on the evaluation of the energy deposition in these experiments in
comparison with our simulation results. The analysis revealed a full compatibility of the FY measurements as long as
the energy deposition in the experimental chamber was accurately determined. This lead us to evaluate an average value
of the FY, preliminary results having been shown in unpublished works \cite{average_arXiv,Thesis} and in several
conferences \cite{UHECR2011,UHECR2012,ICRC2013}. Here, we present an update of our analysis including new relevant
measurements \cite{Dandl,Airfly} and some additional considerations (section \ref{sec:analysis}). The procedure to
obtain our final result of the average FY and the associated uncertainty is given in section \ref{sec:average}.

\section{Upgrades of the Monte Carlo algorithm}
\label{sec:upgrades}

Our MC algorithm was described in detail in previous publications
\cite{PhysLett,AstropartPhys1,NIMA_th,NJP,AstropartPhys2} and therefore only a brief overview is given here. The
algorithm enables the transport of electrons in the range from 100~GeV down to several eV inside an interaction region
of any geometry filled with nitrogen gas at a selected pressure and temperature. An electron trajectory consists of a
succession of ``free-flight'' steps randomly generated from an exponential distribution with average path length equal
to the reciprocal of the product of the gas density and the total cross section. In each step, a discrete random number
is generated to select the kind of interaction: elastic, ionization, excitation (without ionization) or bremsstrahlung,
the probability of each process being proportional to the corresponding cross section. In elastic interactions, the
electron energy does not change but it is scattered in a new random direction according to the differential elastic
cross section. In inelastic interactions, the energy lost by the electron and its angular deflection are calculated
from appropriate data for the N$_2$ molecule. If the inelastic interaction is an excitation, the average amount of 2P
fluorescence light emitted by the excited nitrogen molecule is evaluated as the ratio of the emission cross section and
the total excitation cross section. In the case of ionization, the 1N fluorescence emitted by N$^+_2$ is evaluated
correspondingly, and a secondary electron is ejected with a random kinetic energy and direction according to the
differential ionization cross section and momentum conservation. If the primary electron has enough energy, a K-shell
ionization may occur with a probability equal to the ratio of the K-shell ionization cross section and the total
ionization cross section, resulting in the emission of a X ray too.\footnote{The generation and transport of X rays
were included in \cite{AstropartPhys2}, but not in earlier versions of the algorithm.}

All the individual interactions of both primary and secondary electrons (and X rays) are simulated until they either
leave the considered interaction region or have an energy below 11~eV, which corresponds to the threshold for
fluorescence production. Below this threshold, the electron is assumed to deposit its remaining energy in the medium.
Bremsstrahlung photons are not tracked, since they are assumed to leave the medium without interacting in it. Note
that, unlike many general-purpose MC codes that use the multiple scattering approach, this detailed simulation allows
us to describe the energy deposition and the fluorescence emission at a microscopic scale.

We improved some relevant ingredients of our algorithm since the version presented in \cite{AstropartPhys2}. The
upgrades are described in the following subsections.

\subsection{Energy spectrum of secondary electrons}
\label{ssec:spectrum}

An adequate description of the single-differential ionization cross section $\rmd\sigma_{\rm ion}/\rmd W$ of nitrogen
as a function of the secondary electron energy $W$ and the incident electron energy $T$ is a key ingredient of our MC
algorithm. A simple analytical model of $\rmd\sigma_{\rm ion}/\rmd W$ was presented in \cite{NJP}; however, exchange
effects due to the indistinguishability of the scattered and ejected electrons were neglected. The model has been
upgraded to account for these effects as described in \ref{app:spectrum}.

This upgrade results in a slight increase in the production of high-energy secondary electrons, i.e., $\delta$ rays.
Although they are very scarce, $\delta$ rays carry a significant fraction of the energy lost by the primary particle
and, due to their relatively large ranges, they can leave the field of view of the detector measuring the fluorescence
light in a laboratory experiment. As a consequence of this upgrade, the updated results of the energy deposition in a
finite medium (see section \ref{ssec:energy_deposition}) are somewhat lower compared to previous versions of the
algorithm. On the other hand, our theoretical predictions of the FY (section \ref{ssec:nitrogen_fluorescence}) are
almost unaffected, because $\delta$ rays have an efficiency for fluorescence production similar to that of the primary
electron.

\subsection{Density-effect correction to the K-shell ionization cross section}
\label{ssec:density_effect}

At very high energies ($T\gtrsim1$~GeV), the scattering cross sections have to be corrected for the so-called ``density
effect'', which is due to the polarization of the medium \cite{Sternheimer}. In our algorithm, this correction is
applied to the total ionization cross section as described in \cite{AstropartPhys1}, but it was formerly neglected for
the (partial) K-shell ionization cross section. In the present version of the algorithm, the density-effect correction
has been included in this cross section assuming the same correction term as that used in the total ionization cross
section. This assumption is justified by the fact that differences between the energies of the N$_2$ shells are
negligible compared to the incident electron energies for which this effect is important.

This upgrade somewhat affects the yield of X rays, which are responsible for a significant fraction of the energy
deposition and the fluorescence emission. There is no significant effect on both the total energy loss and the FY,
however, the updated results of the energy deposition in a finite medium for primary electron energies in the GeV range
(see subsection \ref{ssec:energy_deposition}) are slightly lower than those reported in \cite{AstropartPhys2}.

\subsection{Angular distribution of scattered electrons}
\label{ssec:angular_distribution}

The angular deflections of electron trajectories in air or nitrogen are mainly determined by elastic collisions,
although inelastic interactions may also be important. In the present version of our MC algorithm, we have upgraded the
the angular distribution of scattered electrons in elastic collisions as described in \ref{sapp:elasctic}. The
improvement with respect to previous versions is only relevant for low-energy electrons, which have very short ranges
at atmospheric pressure. Therefore, this has no significant impact for the purposes of this paper, although it may be
important at very low-pressure conditions, where an accurate tracking of low-energy secondary electrons is necessary
\cite{Thesis,NIMA_exp}.

In previous versions of the MC algorithm, the angular distribution of scattered electrons in inelastic collisions was
assumed to be the same as the one used for elastic collisions. This crude approximation was justified by the fact that
the final results are only weakly dependent on the details in the tracking of both primary and secondary electrons.
However, we realized that our algorithm slightly overestimated the average path length of primary electrons of low
energies ($T\lesssim1$~MeV), resulting in a small but systematic bias on the calculated energy deposition
\cite{UHECR2012}. To correct this bias, the angular distribution of scattered electrons in ionization collisions has
been revised as described in \ref{sapp:ionizations}. The sampling of the direction of ejection of secondary electrons
is also described in that Appendix.

\section{Updated simulation results}
\label{sec:MC_results}

In general, we are interested in the evaluation of the energy deposition and the fluorescence emission upon the passage
of energetic electrons through a given volume of nitrogen or air. Our MC algorithm calculates both magnitudes as the
sum of all the tiny contributions of the individual inelastic collisions undergone by the primary electrons as well as
by every secondary electron and X ray generated inside the interaction region. The FY is determined as the ratio of the
total fluorescence intensity and energy deposition in the same volume. Only the emissions in two reference bands at
337~nm and 391~nm of the 2P and 1N systems, respectively, are evaluated, since the relative intensities of the
remaining bands in the spectral range of interest are known. The algorithm is for pure nitrogen and it does not include
collisional quenching.

As described in previous works \cite{PhysLett,AstropartPhys1,NJP}, we developed a general-purpose simulation, hereafter
referred to as ``generic simulation'', where the primary electrons are not transported. Instead, they are assumed to
have a single inelastic interaction at a given point and the ejected secondary particles are tracked up to a certain
radial distance $R$. In this way, we obtained both the mean energy deposition and the mean fluorescence emission per
primary inelastic interaction. Updated results are presented in subsections \ref{ssec:energy_deposition} and
\ref{ssec:nitrogen_fluorescence}, respectively.

In addition, realistic simulations including the tracking of primary electrons were carried out for specific
geometries. The corresponding FY results are compared with those of the generic simulation in subsection
\ref{ssec:nitrogen_fluorescence}.

Finally, the theoretical absolute value of the FY for the 337~nm band in air at atmospheric conditions was obtained by
applying the appropriate quenching reduction. The result is shown subsection \ref{ssec:air_fluorescence}.

\subsection{Stopping power and energy deposition}
\label{ssec:energy_deposition}

\begin{figure}[t]
\includegraphics[width=1\linewidth]{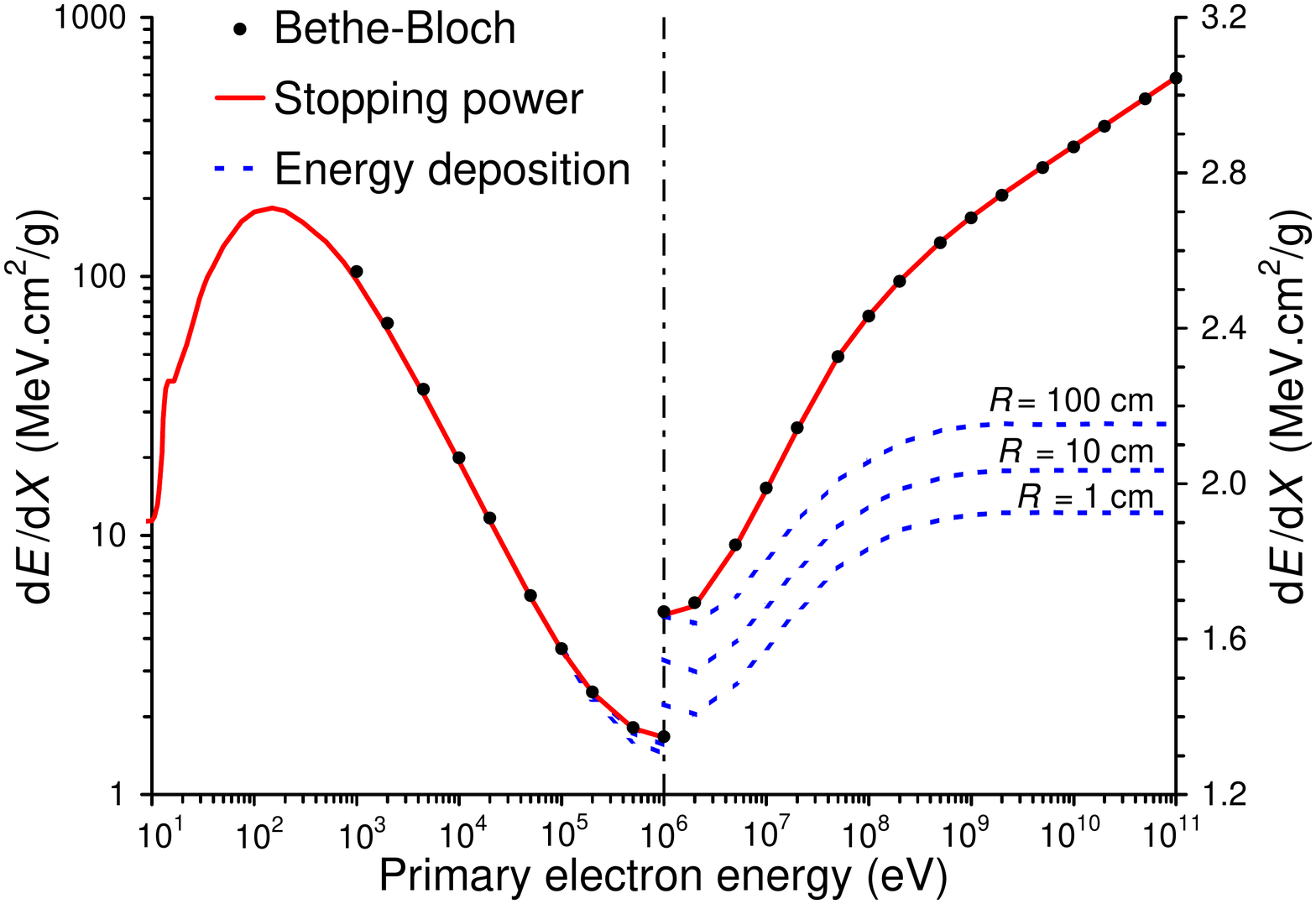}
\caption{%
Updated results of collision energy loss and energy deposition per unit mass thickness of nitrogen at 1.137~mg/cm$^3$ (i.e., 1~atm and 300~K).
Filled circles: Bethe-Bloch stopping power \protect\cite{ICRU}.
Solid line (red): stopping power according to the microscopic treatment implemented in our MC algorithm (see text).
Dotted lines (blue): MC results of energy deposition restricted to spheres of radii $R=1$, 10 and 100~cm around the primary interaction point (see text).
Different vertical scales are used to represent results for energies below 1~MeV (left axis) and above 1~MeV (right axis).%
}
\label{fig:Edep}
\end{figure}

An important requisite of our MC algorithm is to account for the collision stopping power, that is, the mean energy
loss in inelastic collisions per unit mass thickness traversed by an electron. As described in \cite{NJP}, this
parameter is related to the molecular parameters used in the algorithm (i.e., excitation and ionization cross sections,
mean energy of ejected secondary electrons, etc.) by a simple expression and therefore it can be calculated without the
need of a simulation. In figure \ref{fig:Edep}, the result of this calculation is compared with the Bethe-Bloch formula
of the stopping power evaluated from the parameterization given by \cite{ICRU}. Results agree within the uncertainties
of this parameterization, which were estimated by the authors to be around 1\% for energies above 100~keV, but larger
at lower energies because of the omission of the shell correction. As discussed in \cite{ICRU}, this parametrization of
the Bethe-Bloch formula is expected to overestimate the stopping power by about 10\% at 1~keV for low-$Z$ materials and
it is no longer valid at lower energies. Our calculations, which are free from this systematic error, give a stopping
power lower than that of \cite{ICRU} by 3\% at 10~keV and by 8\% at 1~keV, within expectations. This allows us to cover
the vast energy range from 100~GeV down to 10~eV. Note that an adequate description of the slowdown of low-energy
secondary electrons is mandatory for a theoretical evaluation of the FY.

A fraction of the energy lost by the primary electrons is carried away by $\delta$ rays that can escape the interaction
region. Therefore, the energy deposition in the medium is lower than the total energy loss, unless the primary electron
energy is not high enough to produce $\delta$ rays or the interaction region is very large. Results of our generic
simulation for a nitrogen density of 1.137~mg/cm$^3$ (i.e., 1~atm and 300~K) and various $R$ values are shown in figure
\ref{fig:Edep}. Note that the mean energy deposition per primary interaction provided by this simulation has to be
divided by the product of the density and the mean free path of the primary electrons to obtain the differential energy
deposition per unit mass thickness $\rmd E_{\rm dep}/\rmd X$. We obtained that, for a typical cm-sized interaction
region, the energy deposition is about 10\% lower than the energy loss for primary electrons of 1~MeV, and the
difference amounts to more than 25\% in the GeV range. We also found that the energy deposition is only weakly
dependent on the size of the interaction volume (or the nitrogen density). Therefore, a reasonable accuracy in the
energy deposition can be achieved without the need to simulate the fine geometrical details of an experiment.

At usual incident electron energies ($\gtrsim10$~keV), the stopping powers of nitrogen and air are almost identical
(within 1\%) when they are expressed in units of mass thickness.\footnote{\label{fn:stopping_power}The stopping power
expressed in these units is basically proportional to the ratio of the atomic number and the atomic weight of the
material, and this ratio is equal to 1/2 for both nitrogen and oxygen gases, which constitute 99\% of air.} Also, the
fraction of energy carried away by $\delta$ rays that leave the interaction volume is expected to be basically the same
in both gases at given density. Therefore, the above results of energy deposition in nitrogen are also valid for air.

It was shown in \cite{UHECR2012} that our updated results of energy deposition, including their dependencies on energy
and pressure, are in excellent agreement with those obtained with Geant4 \cite{Geant4}. From this comparison, we
estimated an uncertainty of 2\% in the evaluation of the energy deposition at usual conditions of experiments measuring
the FY.

\subsection{Nitrogen-fluorescence yield in the absence of quenching}
\label{ssec:nitrogen_fluorescence}

\begin{figure}[t]
\includegraphics[width=1\linewidth]{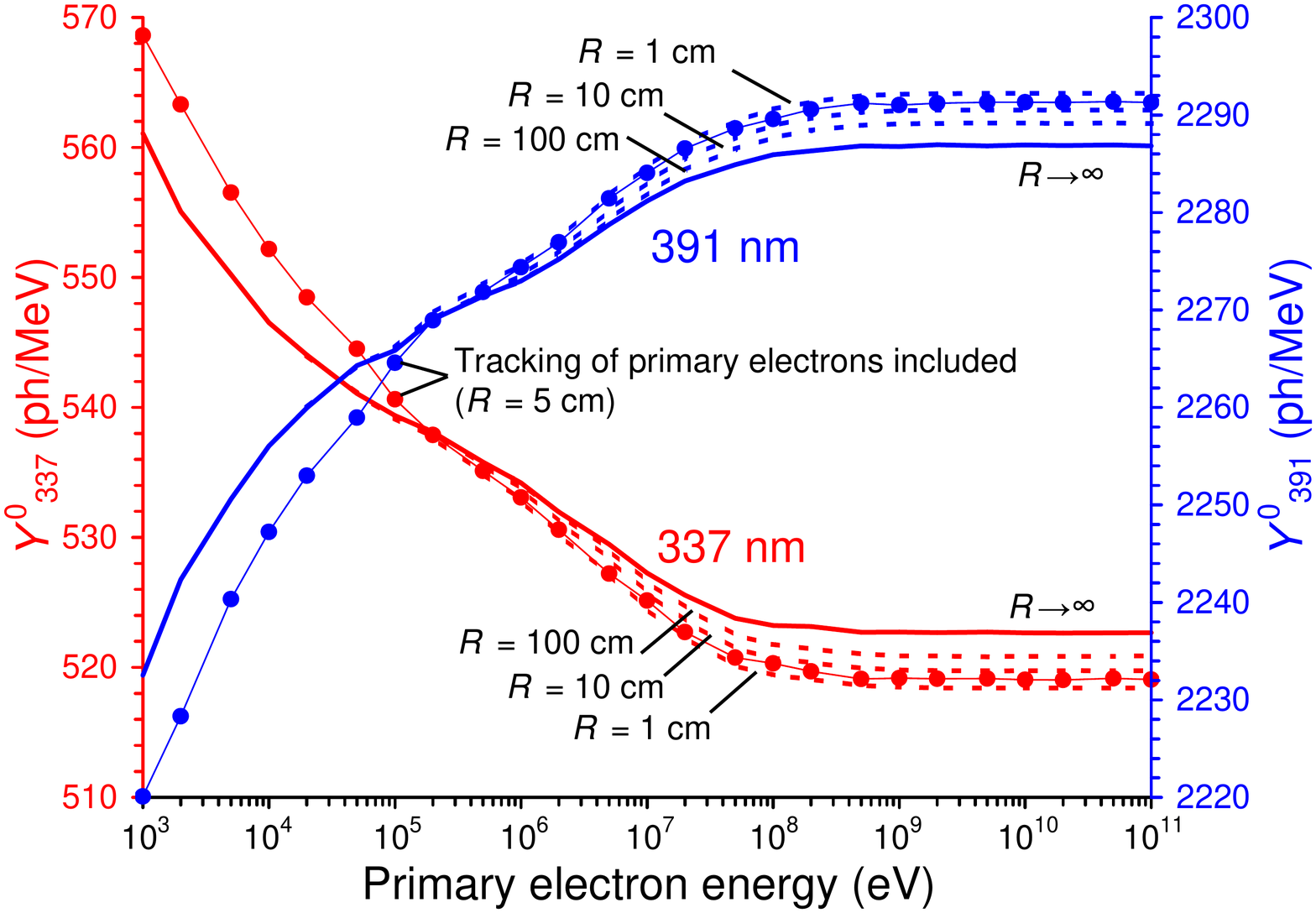}
\caption{%
MC results of the FY in nitrogen at 1.137~mg/cm$^3$ (i.e., 1~atm and 300~K) in the absence of collisional quenching.
Solid thick lines: FY at 337~nm (left axis, in red) and at 391~nm (right axis, in blue) obtained from the generic simulation for an infinite medium.
Dotted lines: FY obtained from the generic simulation restricted to spheres of radii $R=1$, 10 and 100~cm.
Circles connected with lines: FY calculated by tracking both primary and secondary electrons inside a sphere of radius $R=5$~cm.%
}
\label{fig:FY}
\end{figure}

Simulation results of the FY for the bands at 337~nm and 391~nm in nitrogen at 1.137~mg/cm$^3$ in the absence of
collisional quenching are shown in figure \ref{fig:FY}. Our generic simulation predicts that the FY is nearly constant
at usual experimental conditions, that is, the fluorescence intensity is basically proportional to the energy
deposition in a given volume. Variations in the calculated FY are less than 0.2\% at energies above 100~MeV, although a
weak energy dependence is found at lower energies. When the electron energy decreases from 100~MeV down to 1~MeV
(1~keV), the FY for the 337~nm band increases by about 2\% (7\%) and that for the 391~nm band decreases by 0.6\%
(2.5\%).\footnote{The contribution to the energy deposited by an extensive air shower from electrons with energy lower
than 1~MeV is only of about 22\% \protect\cite{Risse}. Therefore, this weak energy dependence of the FY has no relevant
impact on the calibration of the fluorescence telescopes.} Our calculations also show a very slight dependence on the
size of the interaction volume, with variations of less than 1\% for the considered range of $R$ values. These updated
FY results are somewhat greater than those previously presented in \cite{NJP}, where the contribution of X rays to the
fluorescence emission had been neglected.

In practice, the FY is measured as the ratio of the total fluorescence emission and the total energy deposited by the
electron beam in a nitrogen volume of given geometry. The above-mentioned realistic simulations including the tracking
of the primary electrons allow us to evaluate the FY under these conditions. In particular, they account for the fact
that the energy of the primary electrons varies along their trajectories. Results for an electron beam crossing a
sphere of radius $R=5$~cm are shown in the figure as a function of the incident electron energy. At high energies
($\gtrsim1$~MeV), electrons cross the sphere without losing a significant fraction of energy and the FY is fully
consistent with that obtained from the generic simulation. Electrons with energies below 50~keV stop completely in the
sphere, resulting in slightly steeper variation in the FY with energy. For instance, the FY for the 337~nm band at
1~keV is 9\% greater than that at 100~MeV according to this simulation.

Although the fluorescence emission in the 1N system dominates at low pressure, it only represents a small fraction of
the total fluorescence at atmospheric conditions owing to the strong collisional quenching of this band system. The 2P
band at 337~nm is the most intense one at near atmospheric pressure; therefore, it is taken as the reference band for
the analysis of the absolute FY in the following sections.

\subsection{Theoretical value of the absolute air-fluorescence yield}
\label{ssec:air_fluorescence}

The FY for the 337~nm band in air at pressure $P$ is calculated from the following expression:
\begin{equation}
\label{extrapolation}
Y_{337}=\frac{f_{{\rm N}_2}\,Y^0_{337}}{1+P/P'_{337}}\,,
\end{equation}
where $f_{{\rm N}_2}=0.78$ is the molecular fraction of N$_2$ in air, $Y^0_{337}$ is the result of our simulation for
nitrogen in the absence of quenching, and $P'_{337}$ is the characteristic pressure accounting for collisional
quenching in air.

In principle, the FY should be calculated from a simulation for air. However, the above expression is accurate enough
for our purposes. In the first place, both fluorescence intensity and energy deposition per unit path length traveled
by the incident electron are proportional to the amount of generated secondary electrons; thus, the small differences
in the total ionization cross sections of nitrogen and air\footnote{Using data from \cite{NIST_ionization}, we
estimated that the average ionization cross section per air molecule is 3\% larger than that of nitrogen at high
energy. This is mostly due to the differences in the atomic numbers of nitrogen and oxygen (see footnote
\ref{fn:stopping_power}).} are almost exactly canceled out in the FY. Other features, like the slowdown of secondary
electrons in these gases, may also affect the fluorescence production, but the differences in the inelastic cross
sections of N$_2$ and O$_2$ at low energy are comparable to the uncertainties of available measurements
\cite{Itikawa,Itikawa_O2} and therefore they were disregarded in our calculations.

Our MC algorithm predicts $Y^0_{337}\approx520$~ph/MeV for high-energy primary electrons and a reasonably large
interaction volume ($R\gtrsim1$~cm). Using the $P'_{337}$ value from \cite{Airfly_pressure} in expression
(\ref{extrapolation}), we obtained a theoretical value of $Y_{337}=7.9$~ph/MeV at the reference conditions used in the
following sections (i.e., dry air, 800~hPa and 293~K). The uncertainty of this result was estimated to be about 25\%,
which is relatively large because of the contributions of many molecular parameters (e.g., the energy spectrum of
secondary electrons, the stopping power at very low energy and the emission cross section). Efforts are currently being
made to asses the error contributions and ultimately reduce the uncertainty of this theoretical value. Note, however,
that this uncertainty in the absolute value does not apply to the weak dependencies of the FY on energy and the size of
the interaction volume shown in figure \ref{fig:FY}.

Some additional remarks are worth making. Our MC algorithm uses experimental electron-impact emission cross sections
that include possible contributions of cascade from upper lying excited states (see, e.g., \cite{Fons}).\footnote{We
showed in \cite{NJP} that relative intensities of the fluorescence spectrum are proportional to the corresponding
Franck-Condon factors for direct excitations of N$_2$ from the ground state, indicating that cascade effects should be
small.} On the other hand, more complicated excitations channels involving collisions with other molecules
\cite{Grun,Dilecce,Morozov} were not included in our algorithm. Nevertheless, these processes can be neglected in air
because the collisional quenching by O$_2$ dominates. In addition, it has been proved in \cite{Airfly_pressure,Morozov}
that the pressure dependence of the fluorescence intensity is well described by a simple Stern-Volmer law as that
assumed in equation (\ref{extrapolation}), where the effective $P'$ value accounts for all the possible collisional
processes affecting the fluorescence emission. Available measurements of the $P'$ parameter for the band at 337~nm are
in general agreement. Moreover, it was shown in \cite{NJP,Thesis} that the discrepancies between the $P'$ values
reported by M.\ Nagano et al.\ \cite{Nagano} and the Airfly Collaboration \cite{Airfly_pressure} disappear when the
measurements of M.\ Nagano et al.\ are properly corrected to account for $\delta$ rays leaving the experimental
interaction volume.

A recent experimental study carried out by T.\ Dandl et al.\ \cite{Dandl} showed the presence of a long afterglow
following pulsed excitation of pure nitrogen, which could be attributed to recombination processes of N$^+_2$. This
fluorescence contribution is strongly attenuated when the nitrogen gas has traces of impurities of O$_2$ or other
molecular species and it is negligible in air. Therefore, this effect has no impact on both the above theoretical
result and the analysis of air showers detected by fluorescence telescopes.

\section{Critical analysis of absolute measurements of the fluorescence yield}
\label{sec:analysis}

Since the nineties, several measurements of the absolute FY oriented to the cosmic-ray field have been carried out
\cite{Kakimoto,Nagano,Lefeuvre,MACFLY,FLASH,AirLight,Dandl,Airfly}. A summary of these measurements is given in table
\ref{tab:experiments}.

\begin{table*}[t]
\resizebox{1\linewidth}{!}{%
%\centering
\begin{tabular}{l l l l l}
Experiment & Particle & Energy (MeV) & Wavelength interval (nm) & Evaluation of $E_{\rm dep}$ \\
\midrule
F.\ Kakimoto et al.\ \cite{Kakimoto} & e$^-$ & 1.4, 300, 600, 1000              & 337, 300--400 & No             \\
M.\ Nagano et al.\ \cite{Nagano}     & e$^-$ & 0.85                             & 337           & No             \\
G.\ Lefeuvre et al.\ \cite{Lefeuvre} & e$^-$ & 1.1, 1.5                         & 300--430      & No             \\
MACFLY \cite{MACFLY}                 & e$^-$ & 1.5, $2\cdot10^4$, $5\cdot10^4$  & 290--440      & Geant4         \\
FLASH \cite{FLASH}                   & e$^-$ & $2.85\cdot10^4$                  & 300--420      & EGS4           \\
AirLight \cite{AirLight}             & e$^-$ & 0.2--2                           & 337           & Geant4         \\
T.\ Dandl et al.\ \cite{Dandl}       & e$^-$ & $1\cdot10^{-2}$                  & 337           & Not needed$^a$ \\
Airfly \cite{Airfly}                 & p     & $1.2\cdot10^5$                   & 337           & Geant4         \\
\end{tabular}%
}
\begin{flushleft}\footnotesize $^a$Electrons of 10~keV stop completely in an air volume of about 1~mm$^3$.\end{flushleft}%
\caption{%
Measurements of the absolute FY in air available in the literature. The type of particle, energy and wavelength
interval used in each experiment are indicated. The MC code (if any) employed for the evaluation of the energy
deposition
is shown in the last column.%
} \label{tab:experiments}
\end{table*}

Most experiments used electrons either from a $^{90}$Sr radioactive source
\cite{Kakimoto,Nagano,Lefeuvre,MACFLY,AirLight} with an average energy of around 1~MeV, or from accelerators
\cite{Kakimoto,MACFLY,FLASH}, which can provide energies in the GeV range. T.\ Dandl et al.\ \cite{Dandl} employed an
electron gun producing a dc-beam of $\sim10$~keV electrons, and Airfly \cite{Airfly} used 120~GeV protons from the Test
Beam Facility of the Fermi National Accelerator Laboratory.

Narrow-band filters were used in \cite{Kakimoto,Nagano,AirLight,Dandl,Airfly}, providing the FY for the reference 2P
band at 337~nm. On the other hand, some measurements \cite{Kakimoto,Lefeuvre,MACFLY,FLASH} were performed by using
wide-band filters similar to those employed in fluorescence telescopes, which typically collect light in the
300--400~nm spectral range. For these measurements, we made a wavelength normalization to obtain the FY for the 337~nm
band using precise experimental relative intensities from \cite{Airfly_pressure}.\footnote{Theoretical relationships of
the relative intensities given in \cite{NJP}, in full agreement with experimental data, were also used to normalize the
measurements of the MACFLY Collaboration \cite{MACFLY}, which include some weak bands outside the spectral range of
\cite{Airfly_pressure}.} The estimated uncertainty associated with this normalization is 1.7\%, which was
conservatively calculated by treating the systematic uncertainties of the relative intensities as partly correlated, as
suggested by the authors.

In addition, all the results were normalized to common air conditions of 800~hPa and 293~K. If the pressure and
temperature dependence was measured in a given experiment, these data were used for the normalization of the
corresponding FY result. Otherwise, precise quenching data from \cite{Airfly_pressure,Airfly_temperature} were used.
Nevertheless, the use of different possible $P'$ values leads to deviations less than 1\% in the FY values. More
details on this normalization to common wavelength and air conditions are given in \cite{AstropartPhys2}.

\begin{figure}[t!]
%% For the release version
%% \includegraphics[width=1\linewidth]{no_corr_E}
%% \includegraphics[width=1\linewidth]{corr_E}
\centering
\includegraphics[width=0.7\linewidth]{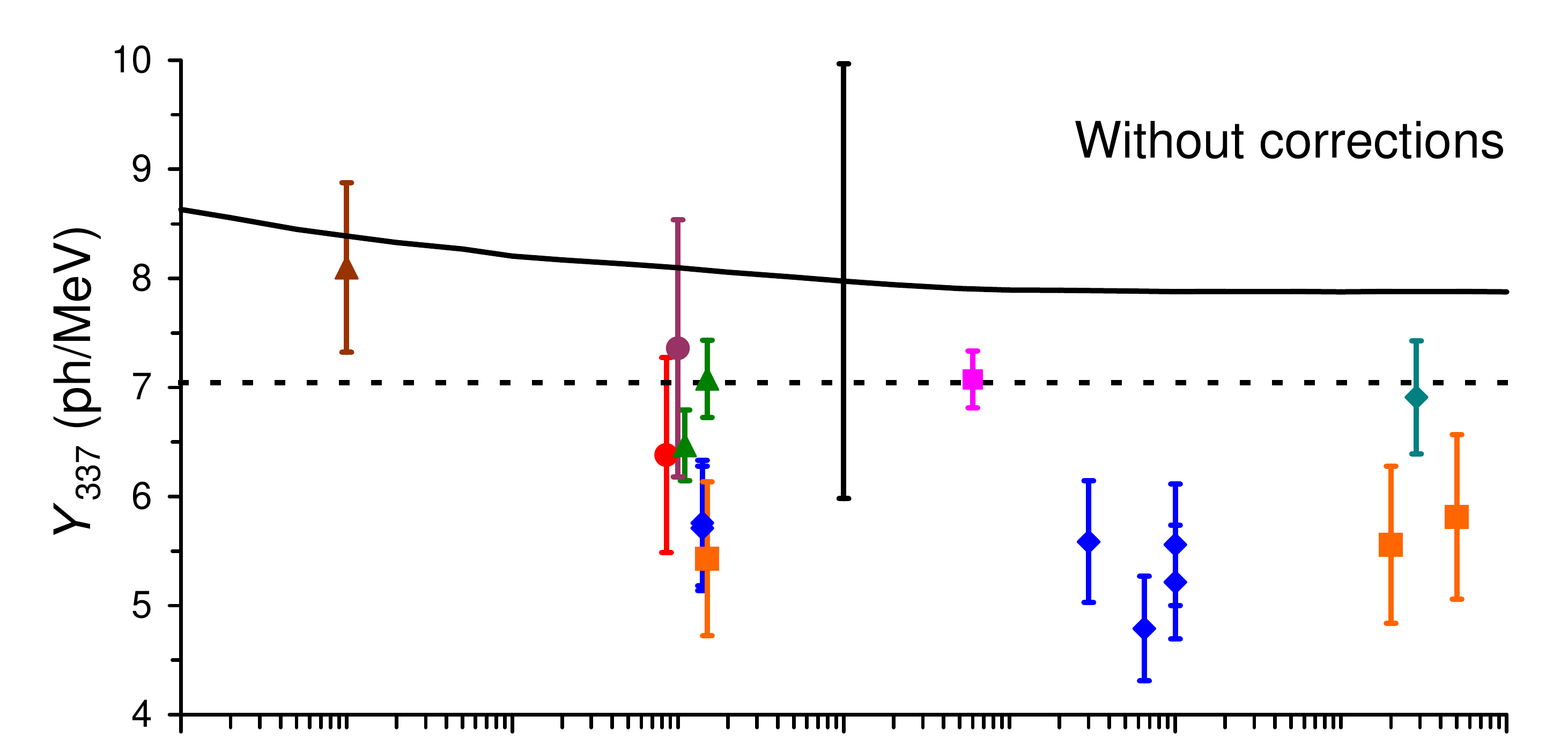}
\includegraphics[width=0.7\linewidth]{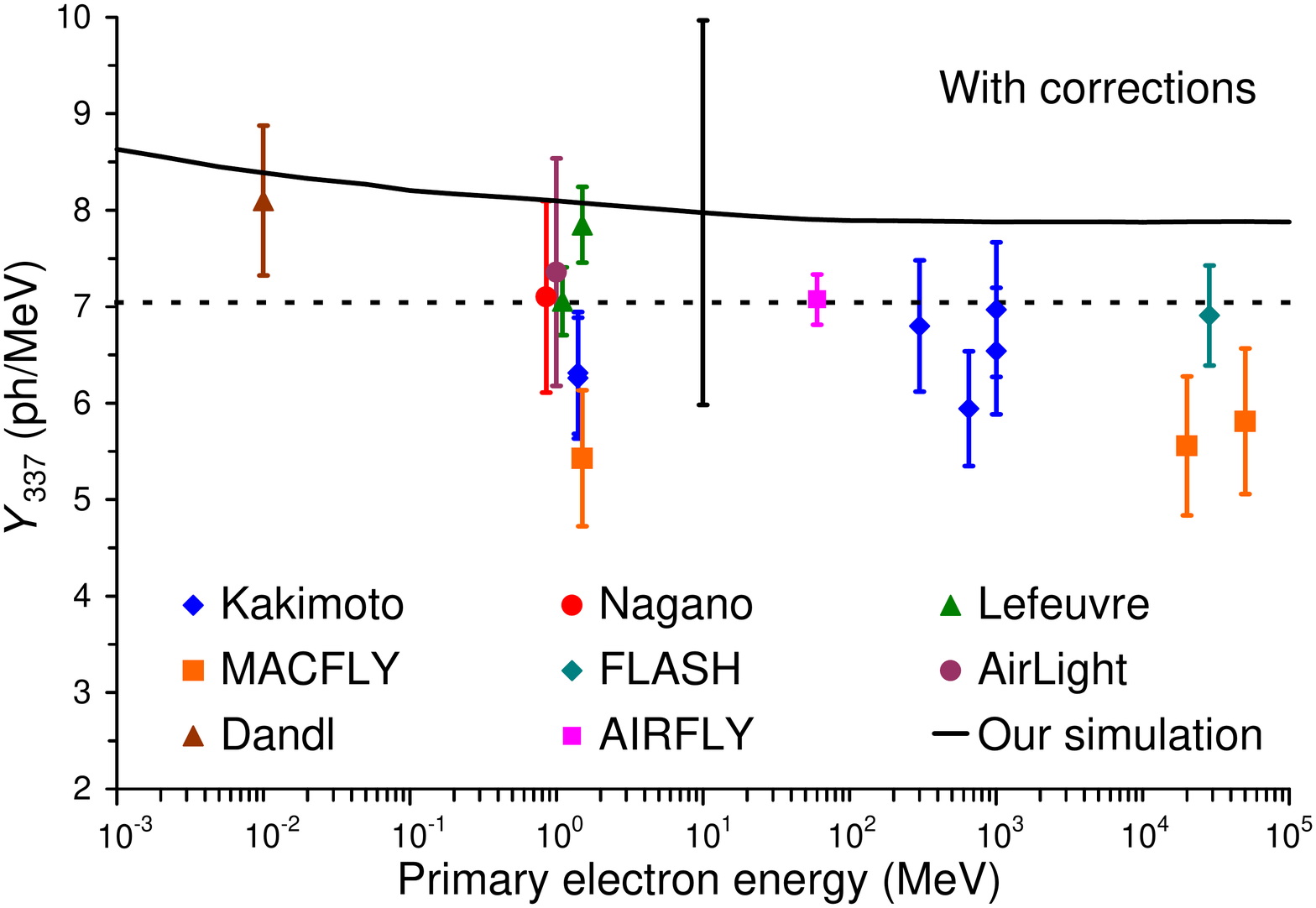}
\caption{%
FY measurements normalized to common conditions (337~nm band in air at 800~hPa and 293~K).
Original results are shown in the top panel. Some measurements were corrected for a bias in the energy deposition in the bottom panel.
The horizontal dashed line at 7.04~ph/MeV represents our average FY determined in section \protect\ref{sec:average}.
The solid line represents the theoretical prediction of the absolute FY from the MC simulation described in \protect\ref{ssec:energy_dependence}
with the corresponding estimated uncertainty (error bar plotted at 10~MeV).%
} \label{fig:comparison_E}
\end{figure}

The normalized FY values are plotted against the incident electron energy in the top panel of figure
\ref{fig:comparison_E}. An electron-equivalent energy of 60~MeV was assigned to the Airfly measurement for 120~GeV
protons (see subsection \ref{ssec:energy_dependence}). A general agreement is found, although discrepancies are larger
than experimental error bars in some cases. The horizontal dashed line at 7.04~ph/MeV represents our average FY
determined in section \ref{sec:average}.

In these experiments, an accurate calculation of the energy deposited in the field of view of the detector is required
to determine the FY. In recent works \cite{MACFLY,FLASH,AirLight,Airfly}, the energy deposition was carefully evaluated
by means of detailed Geant4 \cite{Geant4} or EGS4 \cite{EGS4} simulations. However, in other well-known experiments
\cite{Kakimoto,Nagano,Lefeuvre}, the energy deposition was assumed to be equal to the electron energy loss calculated
from the Bethe-Bloch formula, i.e., they neglected the energy carried away by $\delta$ rays that escape the detector
field of view. Note that this assumption only affects the evaluation of the energy deposition, not the measurement of
the fluorescence intensity. Therefore, the results of these experiments should be considered fully reliable if
appropriate corrections are applied to the energy deposition. These corrections are described in subsection
\ref{ssec:corrections} and they lead to the scenario shown in the bottom panel of figure \ref{fig:comparison_E}.
Additional remarks on the evaluation of the energy deposition are given in that subsection too.

Effects of the possible dependence of the FY on the energy and type of incident particle are discussed in subsection
\ref{ssec:energy_dependence}. Other considerations on the use of synthetic air and pure nitrogen in these experiments
are made in subsection \ref{ssec:gas_composition}.

\subsection{Evaluation of the energy deposition}
\label{ssec:corrections}

As pointed out above, the FY values reported by F.\ Kakimoto et al.\ \cite{Kakimoto}, M.\ Nagano et al.\ \cite{Nagano}
and G.\ Lefeuvre et al.\ \cite{Lefeuvre} have to be corrected to account for a bias in the energy deposition. In
\cite{AstropartPhys2}, we performed a dedicated simulation of the experiment of M.\ Nagano et al.\ including the
geometrical details of their setup and other experimental features. We have redone this detailed simulation
implementing the upgrades described above. The corresponding updated correction on the FY value of this experiment
amounts to +11\%. This result is fully consistent with the predictions from the generic simulation for a sphere of 5~cm
radius that approximately accounts for the interaction region of this experiment.

Following the same procedure of \cite{AstropartPhys2}, we obtained the corrections to the FY results of F.\ Kakimoto et
al.\ and G.\ Lefeuvre et al.\ from the generic simulation for $R=10$~cm and $R=4$~cm, respectively. For the last
experiment, effects of electron scattering by the lead walls of the interaction chamber were also considered in our
calculations (see \cite{AstropartPhys2} for details). As a result, the FY value of F.\ Kakimoto et al.\ at an average
electron energy of 1.4~MeV was increased by 10\%, and those of G.\ Lefeuvre et al.\ at 1.1~MeV and 1.5~MeV were
increased by 9\% and 11\%, respectively. Larger corrections, ranging from +22\% to +25\%, were applied to the FY
results of F.\ Kakimoto et al.\ for incident electrons of 300, 600 and 1000 MeV.\footnote{These results at high energy
were not considered to calculate our average FY, because F.\ Kakimoto et al.\ only used their measurement at 1.4~MeV to
gave their final result.} As illustrated in figure \ref{fig:comparison_E}, our corrections lead to a better agreement
of results.

In addition, we carried out in \cite{AstropartPhys2} detailed simulations of the experiments of MACFLY \cite{MACFLY},
FLASH \cite{FLASH} and AirLight \cite{AirLight}, and they were compared with the simulations performed by the authors
of these experiments. After including the upgrades in our MC algorithm, we found that our simulation results of energy
deposition are in excellent agreement with those of MACFLY using Geant4 as well as with results of other Geant4
simulations that we performed independently \cite{UHECR2012}. On the other hand, the energy deposition determined by
AirLight, also using Geant4, is 5\% lower than our prediction for this experiment. This could be justified by a coding
error in the simulation of AirLight that the authors reported later in \cite{Waldenmaier}. If our simulation results
were used, the FY value of AirLight should be lowered by 5\%. We also found discrepancies between the EGS4 simulation
performed by FLASH and ours, which would lead to a correction of +5\% in their FY value. The origin of these
discrepancies for the FLASH experiment is still unclear, but we observed a possible inconsistency in the treatment of
density effect between both simulations (see \cite{AstropartPhys2,UHECR2011} for details).

Airfly assumed a 2\% contribution to the total uncertainty of the FY due to the determination of the energy deposition
by means of a Geant4 simulation. This uncertainty is consistent with our analysis reported in \cite{UHECR2012}. FLASH
assumed a 1\% uncertainty in their EGS4 simulation. However, the remaining experiments neglected this error
contribution in comparison with more relevant ones.

\subsection{Dependence on the energy and type of the incident particle}
\label{ssec:energy_dependence}

According to our simulations, the FY for the 337~nm band and incident electron energies in the keV range could be up to
9\% greater than that at high energies. In figure \ref{fig:comparison_E}, we show the theoretical FY in air at the
reference conditions obtained from a realistic simulation of an electron beam crossing a sphere of radius $R=5$~cm (see
subsection \ref{ssec:nitrogen_fluorescence}). The experimental data are compatible with this theoretical prediction of
a weak energy dependence of the FY. In particular, the measurement of T.\ Dandl et al.\ \cite{Dandl} for 10~keV
electrons stopping in air is about 15\% greater than those of other measurements at higher energies. The differences
are however comparable to the experimental uncertainties; therefore, these results are not conclusive.

In principle, the FY might also depend on the type of the incident particle. So, the FY measured by Airfly
\cite{Airfly} for 120~GeV protons could be different to that for electrons, which is the one needed for the analysis of
extensive air showers. Nevertheless, as discussed in \ref{app:spectrum}, the energy spectra of secondary electrons
generated by incident electrons and protons of same velocity only differ at the highest $W$ values, and no significant
effect on the FY is expected. To verify this, we implemented in the generic simulation described in section
\ref{sec:MC_results} the energy spectrum of secondary electrons produced by a proton of 120~GeV. The difference between
the resulting FY and the one obtained for 60~MeV electrons is well below 1\%.

\subsection{Effects related to gas composition}
\label{ssec:gas_composition}

The above experiments used synthetic air to ensure experimental conditions free from water vapor and pollutants.
Synthetic air for research purposes is typically a N$_2$-O$_2$ mixture with a nitrogen fraction between 79\% and 80\%,
whereas the composition of atmospheric dry air is 78.08\% of N$_2$, 20.95\% of O$_2$ and 0.93\% of Ar ($<0.1\%$ of
others components). The mass stopping power and the electron trajectories are expected to be insensitive to the small
differences between the compositions of synthetic air-like mixtures and atmospheric air. However, the effect on the
measured FY may be significant, because the fluorescence production is proportional to the nitrogen fraction in the gas
and it also depends on the concentration of the other components through the collisional quenching. Airfly used a
N$_2$-O$_2$ mixture with $(79\pm1)\%$ of nitrogen and, consequently, added a 1\% contribution to the total uncertainty
due to this effect. On the other hand, most experiments neglected this error contribution.

To evaluate the impact of using synthetic air on the FY, we used the following expression:
\begin{equation}
\label{composition}
Y_{\rm air}=Y_{\rm gas}\,\frac{f_{{\rm N}_2}^{\rm air}}{f_{{\rm N}_2}^{\rm gas}}\,
\frac{1+P/P'_{\rm gas}}{1+P/P'_{\rm air}}\,,
\end{equation}
where $Y_{\rm gas}$ is the FY measured in synthetic air at given pressure $P$ and $Y_{\rm air}$ is the one in
atmospheric dry air at same pressure. The characteristic pressure for a given N$_2$-O$_2$ mixture is determined from
the partial $P'$ values for the two components by using quenching data from \cite{Airfly_pressure}. According to this
equation, the measurement of M.\ Nagano et al., who used a mixture of 78.8\% of N$_2$ and 21.2\% of O$_2$, should be
reduced by about 1\% to obtain the FY in dry air. In principle, other results should also be reduced by a similar
amount, but precise data of the composition of the air-like mixtures used by these experiments are not explicitly given
in some cases.

In the experiments of T.\ Dandl et al.\ and Airfly, the absolute measurement of the FY was actually performed in
nitrogen (the fluorescence emission in air is strongly quenched by O$_2$), and the authors converted it into the FY in
air by using the ratio of the fluorescence intensity in nitrogen to that of air obtained in a separate measurement.
However, the intensity ratio measured by T.\ Dandl et al.\ at near atmospheric pressure is about a factor of two higher
than the ratio determined by Airfly in \cite{Airfly}, which agrees with their previous measurement reported in
\cite{Airfly_pressure} as well as with the measurement performed by M.\ Nagano et al. As pointed out in subsection
\ref{ssec:air_fluorescence}, T.\ Dandl et al.\ showed indications of recombination processes that only take place in
very pure nitrogen and that contribute with a long afterglow ($\sim100\,\mu$s) to the total fluorescence. In view of
that, they argued that the discrepancies in the intensity ratios may be due to a higher purity of the nitrogen gas that
they used in their measurements. In principle, this would imply that the FY result of Airfly could be subjected to
unknown uncertainties due to impurities in nitrogen. Nevertheless, the above discrepancies can be justified by the fact
that T.\ Dandl et al.\ measured the continuous light induced by a dc-beam of electrons, whereas both M.\ Nagano et al.\
and Airfly registered fluorescence photons in coincidence with the arrival of single incident particles in a short
interval of few tens of nanosecond, where the contribution from this long afterglow should be negligible. Moreover, in
these coincidence measurements, any small contribution from this afterglow should be subtracted along with the
continuous background. Therefore, we discard any uncertainty due to this effect in the Arfly result.

\section{Determination of an average fluorescence-yield value}
\label{sec:average}

\begin{figure}[t]
\includegraphics[width=\linewidth]{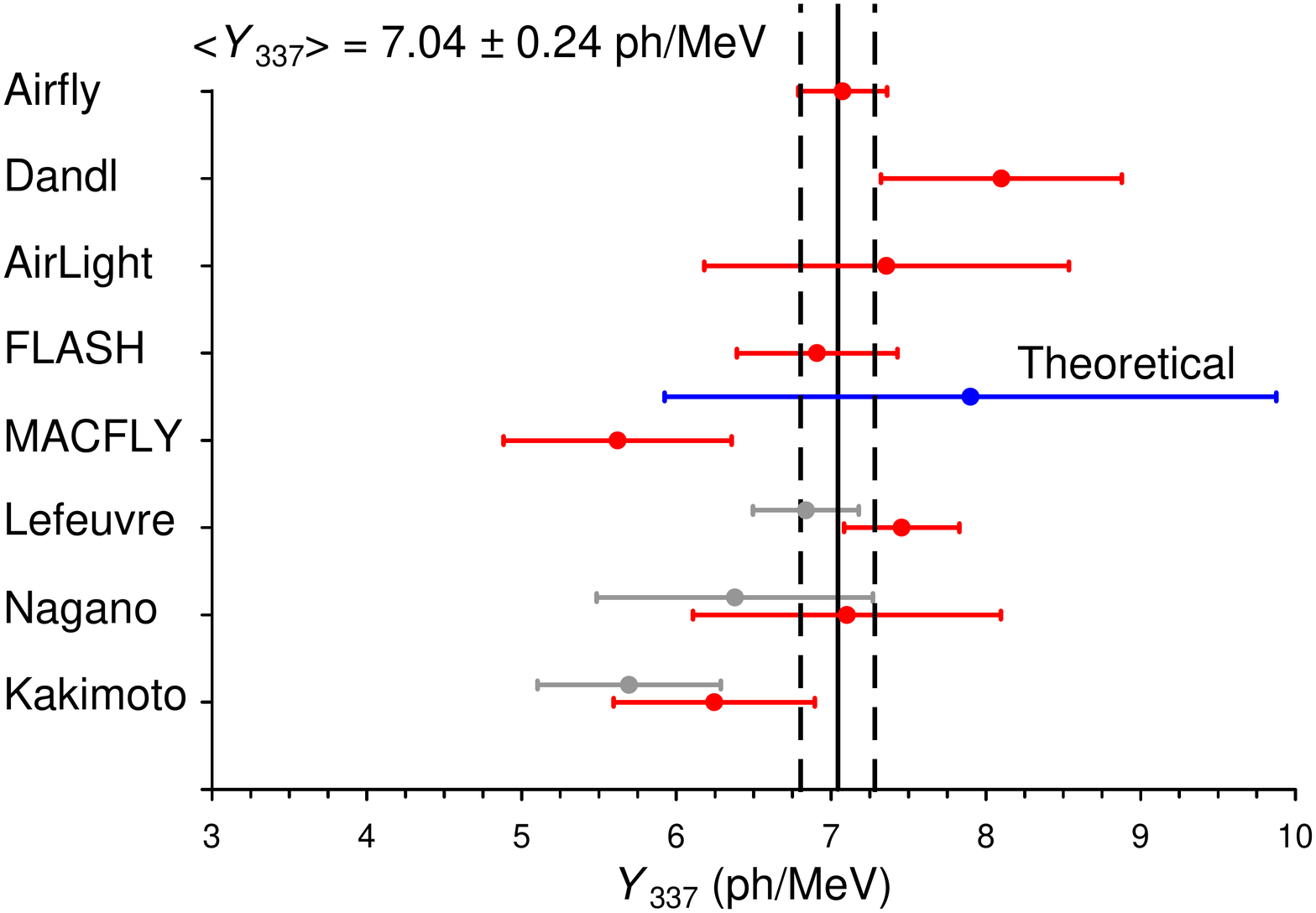}
\caption{%
Graphical representation of the normalized FY values and associated uncertainties (see table \protect\ref{tab:results})).
The original results of F.\ Kakimoto et al.\ \protect\cite{Kakimoto}, M.\ Nagano et al.\ \protect\cite{Nagano} and
G.\ Lefeuvre et al.\ \protect\cite{Lefeuvre} are also shown (grey bars) to illustrate the impact of our corrections.
The solid and dashed vertical lines represent our final result of the average and uncertainty given in the uppermost label
(see subsection \protect\ref{ssec:uncertainty}).
The blue bar represents the theoretical absolute FY predicted by our MC algorithm.%
} \label{fig:comparison_average}
\end{figure}

For the purpose of determining an average FY, we took the final result reported by each experiment, which corresponds
to either a single measurement or a combination of several ones at different conditions (see table
\ref{tab:experiments} and figure \ref{fig:comparison_E}). As described above, all the results were normalized to common
conditions (i.e., the 337~nm band in dry air at 800~hPa and 293~K) and those of F.\ Kakimoto et al., M.\ Nagano et al.\
and G.\ Lefeuvre et al.\ were also corrected for a bias due to the energy deposition. The normalized FY values and the
uncertainties reported by the experiments are shown in table \ref{tab:results} and in figure
\ref{fig:comparison_average}. The original results of F.\ Kakimoto et al., M.\ Nagano et al.\ and G.\ Lefeuvre et al.\
are also shown in the figure (grey bars) to illustrate the effect of our corrections. Our average FY and its
uncertainty interval are represented by vertical solid and dashed lines, respectively. The theoretical FY value
determined in subsection \ref{ssec:air_fluorescence} is shown in the figure too for comparison purposes.

\begin{table}[t]
%% For the release version
%% \resizebox{1\linewidth}{!}{%
\centering
\begin{tabular}{l l l}
Experiment & $Y_{337}$ (ph/MeV) & Uncertainty \\
\midrule
F.\ Kakimoto et al.\ \cite{Kakimoto} & 6.24       & 10\%        \\
M.\ Nagano et al.\ \cite{Nagano}     & 7.10       & 14\%        \\
G.\ Lefeuvre et al.\ \cite{Lefeuvre} & 7.46       & 5\%         \\
MACFLY \cite{MACFLY}                 & 5.62       & 13\%        \\
FLASH \cite{FLASH}                   & 6.91       & 7.5\%       \\
AirLight \cite{AirLight}             & 7.36       & 16\%        \\
T.\ Dandl et al.\ \cite{Dandl}       & 8.10       & 10\%        \\
Airfly \cite{Airfly}                 & 7.07       & 4\%         \\
\end{tabular}%
%}
\caption{%
Normalized FY values and associated uncertainties (337~nm band in dry air at 800~hPa and 293~K). Corrections for
the energy deposition were applied to the results of F.\ Kakimoto et al., M.\ Nagano et al.\ and G.\ Lefeuvre et al.%
} \label{tab:results}
\end{table}

The basic formulations used to compute the average FY and the associated uncertainty are described in subsection
\ref{ssec:checks}. Several consistency checks are also performed in this subsection. The impact on the average of the
several experimental aspects discussed in section \ref{sec:analysis} is evaluated in subsection \ref{ssec:uncertainty}
to determine our final result.

\subsection{Formulations and consistency checks}
\label{ssec:checks}

In principle, we may assume that the quoted experimental uncertainties represent the actual standard deviations of the
corresponding (normal) probability distributions and that there is no correlation between experiments. Under this
assumption, the best estimator of the FY would be
\begin{equation}\label{weighted_mean}
\langle Y \rangle=\frac{\sum_{i=1}^n{Y_i/\sigma_i^2}}{\sum_{i=1}^n 1/\sigma_i^2}\,,
\end{equation}
and the associated uncertainty would be given by
\begin{equation}\label{uncertainty}
\sigma^2_{\langle Y \rangle}=\frac{1}{\sum_{i=1}^n 1/\sigma_i^2}\,,
\end{equation}
where $Y_i$ and $\sigma_i$ are respectively the normalized FY value and the uncertainty of experiment $i$, and $n=8$ is
the number of experiments. The $\chi^2$ statistic divided by the number of degrees of freedom (ndf) is defined in this
case by
\begin{equation}\label{chi2}
\chi^2/{\rm ndf}=\frac{1}{n-1}\sum_{i=1}^n\frac{\left(Y_i-\langle Y\rangle\right)^2}{\sigma_i^2}\,.
\end{equation}

From the data given in table \ref{tab:results}, the above expressions yield an average of 7.06~ph/MeV, a relative
uncertainty of 2.6\% and $\chi^2/{\rm ndf}=1.21$. The probability to obtain a larger $\chi^2$ value is $\alpha=0.29$,
which is an acceptable value indicating that results are consistent with each other. However, the uncertainty of this
average FY is surely not realistic, because the quoted experimental uncertainties may be underestimated in some cases
and other possible effects discussed in section \ref{sec:analysis} can also affect these FY values. Although our final
result was determined by a somewhat different procedure (see subsection \ref{ssec:uncertainty}), expressions
(\ref{weighted_mean}--\ref{chi2}) were used to perform several consistency checks of the data sample as described
below.

We checked that, if a single measurement is removed from the sample, the average value varies by less than $\pm1.7\%$
with respect to that of the full sample. If two any measurements are removed instead, the average varies by less than
$\pm3.9\%$. These deviations in the average are always consistent with the uncertainty calculated from equation
(\ref{uncertainty}) for the corresponding subsample, and the probability $\alpha$ associated with the $\chi^2$ value
takes acceptable values ranging from 0.13 to 0.80. This supports the consistency of results and suggests that the
experimental uncertainties are not correlated.

Our corrections to the measurements of F.\ Kakimoto et al., M.\ Nagano et al.\ and G.\ Lefeuvre et al.\ eliminate a
bias in the average. If these corrections were not included, an average value of 6.82~ph/MeV would be obtained. Note
that this average is lower by an amount of 3.3\%, which is unacceptable at this level of accuracy reached in this
analysis. In addition, the $\chi^2/{\rm ndf}$ statistic of the sample without these corrections is 1.46
($\alpha=0.18$), which is larger than the value of 1.21 ($\alpha=0.29$) obtained for the sample with corrections. This
indicates that the compatibility of results improves when our corrections are applied. Alternatively, if these three
measurements are removed from the sample to avoid any correction, an average of 7.00~ph/MeV is obtained with an
associated uncertainty of 3.2\% and $\chi^2/{\rm ndf}=1.42$ ($\alpha=0.22$). This result is fully compatible within
uncertainties with the average for the whole sample with corrections.

We note that the above value of 7.06~ph/MeV and the result of Airfly are almost identical. This is partly due to the
high precision of the Airfly measurement, which dominates the average. Nevertheless, even though this measurement were
excluded from the data sample, a very close average of 7.04~ph/MeV would be obtained with an associated uncertainty as
low as 3.3\% and $\chi^2/{\rm ndf}=1.41$ ($\alpha=0.21$). Therefore, we can conclude that the result of Airfly is in
full agreement with previous measurements and an average of the whole data sample provides a better precision.

\subsection{Final result}
\label{ssec:uncertainty}

In a preliminary analysis \cite{ICRC2013}, the FY result of 7.06~ph/MeV obtained directly by applying equation
(\ref{weighted_mean}) was taken as the most likely mean value, and a simple procedure to evaluate a conservative
uncertainty was followed. Here, we have employed a more rigorous method that leads to a very similar result. Several
aspects previously discussed in section \ref{sec:analysis} were included in the present analysis:

1.\ We estimated an uncertainty of $\pm2\%$ in the evaluation of the energy deposition in all measurements except for
that of T.\ Dandl et al.

2.\ The wavelength normalization of the FY results of F.\ Kakimoto et al., G.\ Lefeuvre et al., MACFLY and FLASH has an
associated error contribution of $\pm1.7\%$.

3.\ The normalization to common pressure and temperature implies an error contribution that is conservatively estimated
to be of $\pm1\%$.

4.\ The use of synthetic air in these experiments can lead to FY values systematically larger by about 1\%.

5.\ According to our simulation, the FY result of FLASH should be increased by 5\% and that of AirLight should be
decreased by 5\%.

6.\ If measurements are normalized to an electron energy of 100~MeV according to the theoretical relative energy
dependence predicted by our simulation, the FY results at around 1~MeV of F.\ Kakimoto et al., M.\ Nagano et al., G.\
Lefeuvre et al.\ and AirLight should be decreased by 2--3\% and that of T.\ Dandl et al.\ at 10~keV should be decreased
by 6\%.

The uncertainties mentioned in points 1--3 should show some degree of correlation between experiments. Note, for
instance, that the corrections that we applied to the FY values of F.\ Kakimoto et al., M.\ Nagano et al.\ and G.\
Lefeuvre et al.\ are all evaluated with the same MC algorithm, and that measurements are typically carried out at an
air pressure higher than 800~hPa. To be conservative, we made the extreme assumption that these systematic
uncertainties are fully correlated.

For the points 4--6, instead of applying the proposed corrections, we conservatively treated them as additional
uncertainty components (either positive or negative). For instance, point 5 was accounted for by considering an
uncertainty components of +5\% and $-5\%$ on the FY results of FLASH and AirLight, respectively. For the points 4 and
6, we assumed that the corresponding uncertainties are fully correlated between experiments (e.g., the systematic
deviation of 1\% in the FY due to the use of synthetic air affects either all of the measurements or none of them).
Note that this treatment involves asymmetric uncertainties.

The procedure was as follows. First, the average and the associated uncertainty were calculated from equations
(\ref{weighted_mean}) and (\ref{uncertainty}) using only the uncorrelated part of the experimental uncertainties. In
particular, the error contributions due to the energy deposition and the air composition included in the total
uncertainties reported by FLASH and Airfly were removed in this step. The resulting average is still 7.06~ph/MeV and
the corresponding (partial) uncertainty is $\pm2.4\%$. Then, for each error source 1--6, we recalculated the average
after modifying the pertinent FY values in their full uncertainty interval. The corresponding relative shifts of the
average with respect to the above value of 7.06~ph/MeV are listed in table \ref{tab:uncertainties}. All the relative
shifts of same sign were added quadratically. As a result, we obtained an asymmetric systematic uncertainty interval of
$^{+2.3\%}_{-2.7\%}$ around 7.06~ph/MeV, which was converted into a symmetric interval of $\pm2.5\%$ around
7.04~ph/MeV. Finally, this systematic uncertainty was added quadratically to the one previously calculated from the
uncorrelated uncertainties, resulting in a total uncertainty of 3.5\%. Therefore, our final result is $\langle
Y_{337}\rangle=7.04\pm0.24$~ph/MeV in dry air at 800~hPa and 293~K (see figure \ref{fig:comparison_average}).

\begin{table}[t]
%% For the release version
%% \resizebox{1\linewidth}{!}{%
\centering
\begin{tabular}{l l l}
Error source & $-\Delta\langle Y_{337}\rangle$ & $+\Delta\langle Y_{337}\rangle$ \\
\midrule
Energy deposition                                & 1.9\% & 1.9\% \\
Wavelength normalization                         & 0.7\% & 0.7\% \\
Pressure and temperature normalization           & 1\%   & 1\%   \\
Air composition                                  & 1\%   & -     \\
Corrections to the results of FLASH and AirLight & 0.1\% & 0.5\% \\
Energy normalization                             & 1.1\% & -     \\
\midrule
Quadratic sum                                    & 2.7\% & 2.3\% \\
\end{tabular}%
%}
\caption{%
Impact on the average of the error sources indicated in the points 1--6 of subsection \protect\ref{ssec:uncertainty}.
The negative and positive relative shifts with respect to the value of 7.06~ph/MeV are shown in columns 2 and 3.
The quadratic sums of these error contributions are given in the last row.%
} \label{tab:uncertainties}
\end{table}

\section{Conclusions}
\label{sec:conclusions}

The problem of the absolute value of the air-fluorescence yield was studied in depth by means of both a detailed MC
simulation and a critical analysis of available measurements of the FY.

We carried out a theoretical evaluation of the absolute FY based on an upgraded MC algorithm that enables the
calculation of the energy deposition and the fluorescence emission upon the passage of an energetic electron through a
nitrogen volume of given geometry. The algorithm simulates all the individual interactions of both the incident
electrons and the secondary particles (i.e., electrons and X rays) generated in the medium. Special care was paid in
describing correctly the production of low-energy secondary electrons, which are the main responsible for the
fluorescence emission. We obtained that the FY is independent of the incident electron energy (within 3\%) for energies
above 1~MeV, whereas a weak energy dependence was found in the keV region. From our updated simulation results, we
obtained a theoretical absolute FY value of $7.9\pm2.0$~ph/MeV for the band at 337~nm in dry air at 800~hPa and 293~K,
which is in good agreement with experimental values.

We analyzed the available measurements of the absolute FY considering many aspects. In particular, the evaluation of
the energy deposited in the experimental interaction chamber was proved to be crucial for an accurate determination of
the FY. Our simulation results of energy deposition were used to apply corrections to the FY results of some
experiments that neglected the contribution of $\delta$ rays that leave the field of view of the detector measuring the
fluorescence light. Other possible effects (e.g., the weak energy dependence predicted by our MC algorithm and the
differences between the atmospheric air and the synthetic air-like mixtures used in these experiments) were also shown
to be significant. As a final result of this analysis, we obtained an average absolute FY value of $7.04\pm0.24$~ph/MeV
at the reference conditions (i.e., 337~nm band, dry air at 800~hPa and 293~K). This result is very similar to the one
recently reported by the Airfly Collaboration \cite{Airfly}, but with a slight improvement in precision as a
consequence of the consistency of all the FY measurements.

\section*{Acknowledgement}
\label{acknowledgement}

This work was supported by MINECO (FPA2009-07772, FPA2012-39489-C04-02) and CONSOLIDER CPAN CSD2007-42. We thank our
colleagues of the Auger Collaboration for fruitful discussions and comments on this work.

\appendix

\section{Energy spectrum of secondary electrons} \label{app:spectrum}

In the present upgraded version of the MC algorithm, the energy spectrum of secondary electrons generated by low
incident electrons ($T<710$~eV) is described by
\begin{equation}\label{Opal2}
\frac{\rmd\sigma_{\rm ion}}{\rmd W}=K_{\rm low}\left[\frac{1}{w^2+W^2}+\frac{1}{w^2+(T-I-W)^2}\right]\,.
\end{equation}
This is the same empirical formula originally derived by \cite{Opal} that was used in previous versions of our
algorithm, but with the addition of an exchange term that is only relevant near the cutoff energy of the spectrum.
Here, $K_{\rm low}$ is a normalization factor (see below), $w$ is an adjustable parameter that is set equal to 11.4~eV
for N$_2$, and $I=15.6$~eV is the ionization potential. The fastest of the two free electrons after an ionization
collision is customarily defined as the primary (incident) one; therefore, the maximum energy of the secondary electron
is $(T-I)/2$.\footnote{In our MC algorithm, an average excitation energy of $\langle E^{\rm ion}_{\rm
exc}\rangle=1.3$~eV is assumed to be transferred to the molecule in every ionization event \protect\cite{NJP}.
Consequently, the parameter $I$ is in practice replaced with $I+\langle E^{\rm ion}_{\rm exc}\rangle=16.9$~eV, and the
maximum energy of secondary electrons is reduced accordingly for a correct energy balance.}

For $T>710$~eV, we followed the same procedure of \cite{NJP}, but incorporating exchange terms. An analytical
expression of $\rmd\sigma_{\rm ion}/\rmd W$ was constructed so that it approaches expression (\ref{Opal2}) at low $W$
and reproduces the exact M{\o}ller formula for electron-electron scattering cross section \cite{Landau} in the high-$W$
limit. Also, this expression was adapted to be finite at $W=0$ and symmetric under the change $W\rightarrow T-I-W$. The
result is
\begin{align}
\frac{\rmd\sigma_{\rm ion}}{\rmd W}=&\frac{4\pi\,Z\,a^2_0\,\alpha^2\,R}{\beta^2}
\left\{\left[1+K_{\rm high}\,\exp\left(-\frac{W}{W_{\rm t}}\right)\right]\right. \nonumber\\
&\left[\frac{1}{w^2+W^2}+\frac{1}{w^2+(T-I-W)^2}\right] \nonumber\\
&\left.-\frac{2\gamma-1}{\gamma^2\,(W+I/2)(T-W-I/2)}+\frac{1}{(\gamma mc^2)^2}\right\}\,.
\label{ext_Opal2}
\end{align}
where $Z=14$ for molecular nitrogen, $a_0$ is the Bohr radius, $\alpha$ is the fine-structure constant, $R$ is the
Rydberg energy, $\beta$ is the ratio of the incident electron speed to the speed of light $c$, $\gamma$ is the Lorentz
factor, and $m$ is the electron rest mass. The transition energy $W_{\rm t}=71$~eV was tuned so that the Bethe-Bloch
stopping power is correctly reproduced (see section \ref{ssec:energy_deposition}), and the $T$ value at which the
transition from (\ref{Opal2}) to (\ref{ext_Opal2}) takes place was chosen to provide a smooth join between both
expressions.

The factors $K_{\rm low}$ and $K_{\rm high}$ in the above equations are determined by the normalization condition
\begin{equation}\label{normalizacion}
\sigma_{\rm ion}=\int \rmd W \frac{\rmd\sigma_{\rm ion}}{\rmd W}\,,
\end{equation}
where $\sigma_{\rm ion}$ is the total ionization cross section, which includes the density-effect correction at high
energy \cite{AstropartPhys1,Thesis}.

\begin{figure}[t]
\includegraphics[width=1\linewidth]{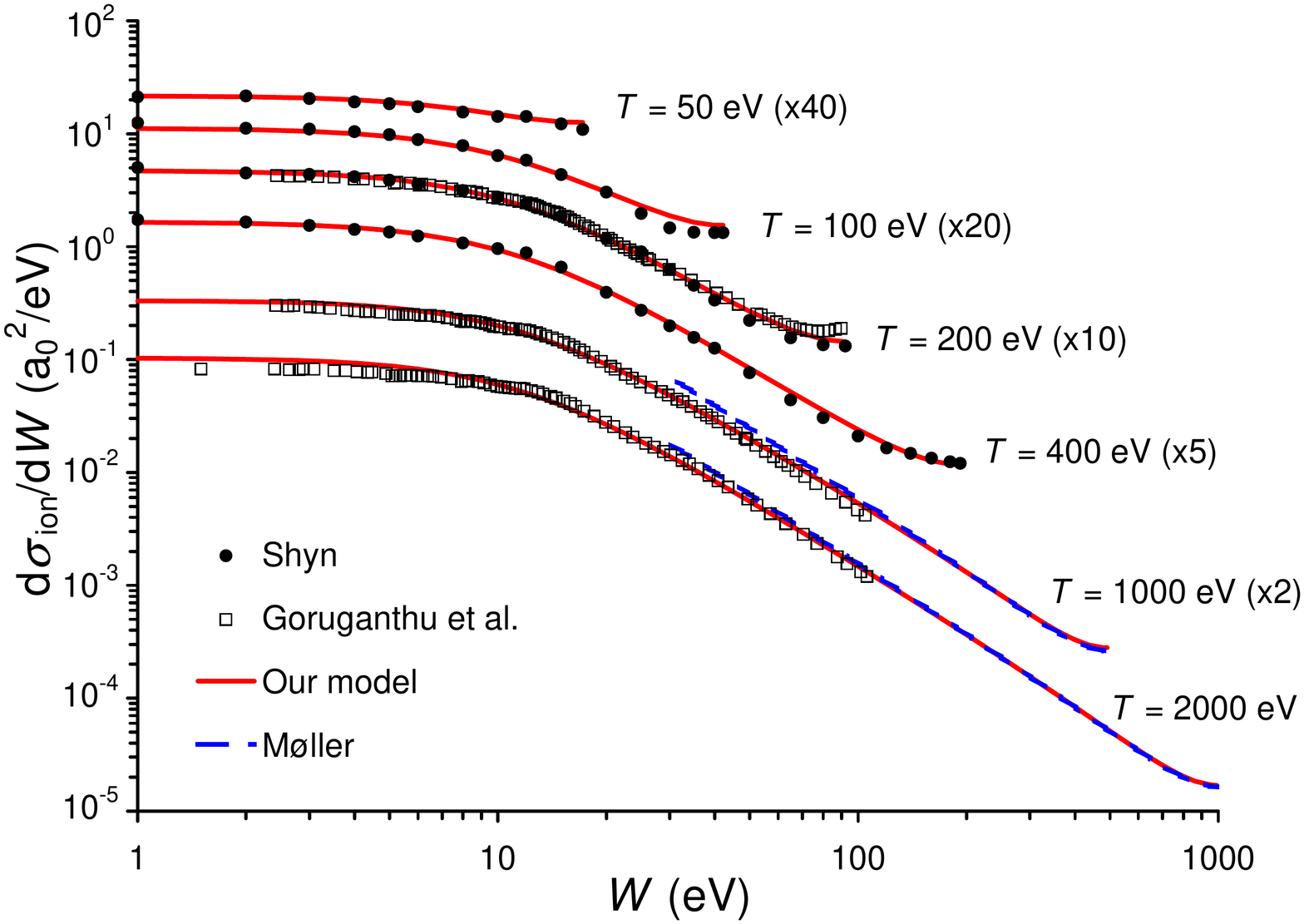}
\caption{%
Comparison of our model of the energy spectrum of secondary electrons with experimental data.
Filled circles: experimental data of \protect\cite{Shyn}.
Empty squares: experimental data of \protect\cite{Goruganthu}.
Solid lines (red): results of expressions (\protect\ref{Opal2}) and (\protect\ref{ext_Opal2}) in their corresponding $T$ domains.
Dashed lines (blue): M{\o}ller cross section.%
}
\label{fig:dist_sec}
\end{figure}

As shown in figure \ref{fig:dist_sec}, results of equations (\ref{Opal2}) and (\ref{ext_Opal2}) are in good agreement
with available experimental data \cite{Shyn,Goruganthu} in the corresponding $T$ domains, and they reproduce correctly
the flatter behavior of the spectra near the cutoff energy as a consequence of exchange effects.

Our model assumes an structureless form (\ref{Opal2}) at low $W$ that does not account for the fine details in the
shape of the experimental spectra. This may be very relevant for a theoretical determination of the FY due to the
strong energy dependence of the emission cross sections in the 10--100~eV range \cite{Itikawa}. We are studying the use
of more sophisticated models of the energy distribution of secondary electrons (see, e.g., \cite{Kim}). Preliminary
tests have indicated us that our approximation may imply a systematic error of up to 10\% on the absolute FY value
given in subsection \ref{ssec:air_fluorescence}, for which we estimated a total estimated uncertainty of 25\%. In
contrast, ignoring these details in the low-$W$ portion of the spectrum pose no impact on our results of energy
deposition (subsection \ref{ssec:energy_deposition}), because the accuracy of this parameter is basically limited by
the uncertainties in the generation and transport of $\delta$ rays.

Analytical expressions for the energy spectrum of secondary electrons generated by incident particles other than
electrons can be derived in a similar way. In fast collisions for which the first Born approximation is valid, the
scattering cross section is proportional to the square of the projectile charge, that is, it is independent of the sign
of charge. In addition, an important conclusion of the Bethe theory is that the cross section is basically independent
of the projectile mass. As a consequence, both the integrated and differential ionization cross sections are almost
identical for fast electrons and protons of same velocity. Differences in the energy spectra of secondary electrons
generated by these two particles only arise near the cutoff energy, which is obviously dependent on the projectile
mass\footnote{Conservation of energy and momentum restricts the maximum energy transferred by a proton to a free
electron at rest to $4T$, where $T$ is the kinetic energy of an electron with the same velocity as the proton.} and
where exchange effects (present in electron-electron scattering, but not in proton-electron scattering) are important.

\section{Angular distributions} \label{app:kinematics}

\subsection{Ionization collisions}\label{sapp:ionizations}

Let us assume a binary collision of an incident electron of kinetic energy $T$ with a target electron initially at rest
in the laboratory frame. After the collision, the incident electron is scattered with an energy $T'$ at a polar angle
$\theta$ and an azimuthal angle $\phi$. The target electron recoils with an energy $W$ at polar angle $\theta_{\rm s}$
and azimuthal angle $\phi_{\rm s}=\phi+\pi$. Due to the axial symmetry about the direction of incidence, $\phi$ is a
uniform random variable in the interval $(0,\,2\pi)$. Once the scattering plane is defined, the kinematics of the
collision is entirely determined by energy and momentum conservation.

On the other hand, in an ionization collision, the incident electron also transfers energy and momentum to the atom or
molecule, and orbital electrons are not at rest, but move with a certain momentum distribution. In our MC algorithm,
the energy transfer $T-T'$ is assumed to be $W+I$ in outer-shell ionizations and $W+I_{\rm K}$ in K-shell ionizations,
where $I_{\rm K}=410$~eV is the ionization potential of the K shell of nitrogen. The directions of motion of the
scattered and ejected electrons are determined by assuming a binary collision, that is, neglecting both the momentum
transferred to the molecule and the initial momentum of the target electron. In principle, this approximation is only
valid in close collisions with large momentum transfer (see, e.g., \cite{Fernandez-Varea}). Nevertheless, we can use
it, with no lack of accuracy for our purposes, also in distant ionization collisions of high-energy electrons, where
the momentum transfer is small and therefore $\theta\approx0$. The angular distributions of scattered and ejected
electrons of low energy cannot be described by this procedure; however, we applied it for all energies because the
tracking of low-energy electrons has no relevant impact on the final results at near atmospheric conditions.

As an improvement only appreciable at low-pressure conditions, an angular straggling is introduced in the generation of
secondary electrons. Experimental data of the angular distribution \cite{Goruganthu} can be approximately described by
a Lorentzian function, where the mean value $\langle\theta_{\rm s}\rangle$ coincides with the angle corresponding to a
binary collision and the width of the distribution decreases with $W$. From a fit to experimental data, we found that
$\theta$ can be sampled from
\begin{equation}
\label{angle_straggling}
\theta_{\rm s}=\langle\theta_{\rm s}\rangle+\frac{0.98}{1+W/65}\tan\left[\frac{\pi}{2}(2r-1)\right]\,
\end{equation}
where $W$ is given in eV and $r$ is a random number in the interval (0, 1). Note that $\theta=\langle\theta_{\rm
s}\rangle$ in the high-$W$ limit, as expected.

\subsection{Elastic collisions} \label{sapp:elasctic}

Theoretical calculations of both the integral and differential elastic cross sections of nitrogen from
\cite{Blanco,Roldan} are used in our MC algorithm. Numerical data of the total elastic cross section $\sigma_{\rm el}$
are included in a database file, and values at given energy are obtained by log-log interpolation. For numerical
purposes, $\theta$ values are efficiently sampled by using an approximated parameterization of the normalized angular
distribution
\begin{equation}
\label{P_theta}
P(\theta)=\frac{\sin\,\theta}{2\pi\sigma_{\rm el}}\frac{\rmd\sigma_{\rm el}}{\rmd \Omega}\,.
\end{equation}
Again, $\phi$ is distributed uniformly.

In the present version of our MC algorithm, the model of $P(\theta)$ was revised to achieve a better agreement with
these theoretical calculations. The following ansatz was assumed:
\begin{align}
P(\theta)=&(1-K)\frac{(1+a_1)\sin\,\theta}{2\left[1+a_1\sin^2(\theta/2))\right]^2} \nonumber\\
&+K\frac{(1+a_2)\sin\,\theta}{2\left[1+a_2\cos^2(\theta/2))\right]^2}\,.
\label{model_P_theta}
\end{align}
The first (dominant) term of the right-hand side of this equation has the form corresponding to electron scattering by
a Yukawa potential in the Born approximation, and the second (symmetric) term was introduced to account for the
relatively large probability of backscattering at low energies. Then, we searched for suitable expressions for the
dimensionless parameters $K$, $a_1$ and $a_2$ that fit theoretical data from \cite{Blanco,Roldan}.

\vspace{0.3cm} \noindent
For $T<40$~eV, we found ($T$ expressed in eV):
\begin{align}
K=&0.4458(1-\rme^{-T/33.98}) \nonumber\\
a_1=&0.4062\frac{T}{6.384}\left(1+\frac{T}{6.384}\right) \nonumber\\
a_2=&9.658\left(1+\frac{T}{354.4}\right)\,.
 \label{0_40}
\end{align}
For 40~eV~$\leq T<700$~eV:
\begin{align}
K=&0.7876\left(1+\frac{T}{23.22}\right) \nonumber\\
a_1=&1.475T^{0.7018} \nonumber\\
a_2=&9.210\rme^{-T/135.3}\,.
 \label{40_700}
\end{align}
And for $T\geq700$~eV:
\begin{align}
K=&0 \nonumber\\
a_1=&18.1\left(1+\frac{T^2/2mc^2+T}{121.7}\right) \nonumber\\
a_2=&0\,.
 \label{700_inf}
\end{align}

\end{document}